\documentclass[prl,twocolumn,superscriptaddress,showpacs]{revtex4}

\usepackage{graphics}           
\usepackage{bm}                 
\usepackage{amsmath}            
\usepackage{amsfonts}
\usepackage{amssymb}

\begin{document}

\title{Typical entanglement in multi-qubit systems}

\author{Vivien M. Kendon}
\email{Viv.Kendon@ic.ac.uk}
\affiliation{Optics Section, Blackett Laboratory, Imperial College,
             London, SW7 2BW, United Kingdom.}
\author{Kae Nemoto}
\affiliation{School of Informatics, Dean Street, University of Wales,
	     Bangor, LL57 1UT, United Kingdom.}
\author{William J Munro}
\affiliation{Hewlett-Packard Laboratories, Filton Road, Stoke Gifford,
	     Bristol, BS34 8QZ, United Kingdom.}

\date{5 June 2001, revised 5 November 2001}

\begin{abstract}
Quantum entanglement and its paradoxical properties hold the key to an
information processing revolution.
Much attention has focused recently on the challenging problem of
characterizing entanglement.
Entanglement for a two qubit system is reasonably well understood,
however, the nature and properties of multiple qubit systems are
largely unexplored.
Motivated by the importance of such systems in quantum computing,
we show that typical pure states of $N$ qubits are highly entangled
but have decreasing amounts of pairwise entanglement
(measured using the Wootter's concurrence formula) as $N$ increases.
Above six qubits very few states have any pairwise entanglement,
and generally, for a typical pure state of $N$ qubits there is a sharp
cut-off where its subsystems of size $m$ become PPT
(positive partial transpose i.e., separable or only bound entangled)
around $N \gtrsim 2m + 3$, based on numerical analysis up to $N=13$.
\end{abstract}

\pacs{03.67.-a, 03.65.Ud}


\maketitle


Quantum entanglement is a key prediction of quantum mechanics
and is generally thought to be one of the crucial resources required in quantum
information processing.
Known quantum information applications include
quantum computation \cite{divincenzo95a,vedral98b}, quantum
communication \cite{schumacher96a},
quantum cryptography \cite{ekert91a,jennewein00a,naik00a,tittel00a},
and quantum teleportation \cite{bennett93a,bouwmeester97a,boschi98a}.
The properties of entangled states potentially used in these
applications are still poorly understood.
General two qubit entangled states have been well
characterized, and a number of analytical measures of entanglement are
known \cite{wootters97a,zyczkowski98a}.  However, for $N$ qubit systems
(with $N>2$), few such measures can be calculated even for pure states.
Despite these difficulties, arrays of qubits have been the focus of
recent attention \cite{wootters00a,oconnor00a,koashi00a,gunlycke01a}, e.g.,
Raussendorf and Briegel \cite{raussendorf01a}
propose an array of highly entangled qubits for quantum computation.

In this letter we consider the average, or typical
properties of $N$ qubit pure states.  A subsystem of an
entangled pure state is in general a mixed state, thus we also
consider the entanglement of mixed states derived from
these pure states.
Currently, a variety of measures are known for quantifying the
degree of entanglement, including
the entanglement of distillation \cite{bennett96b},
the relative entropy of entanglement \cite{vedral97b},
the entanglement of formation \cite{wootters97a,bennett96b}
and the negativity \cite{zyczkowski98a}.
Using $|0\rangle$ and $|1\rangle$ for the two eigenstates of a
spin$-\frac{1}{2}$ system or equivalent that encodes a qubit,
and $|00\rangle$ as a shorthand for $|0\rangle_A|0\rangle_B$,
a state of two qubits $A$ and $B$,
the four Bell states
$(|00\rangle \pm |11\rangle)/\sqrt{2}$ and
$(|01\rangle \pm |10\rangle)/\sqrt{2}$
have the maximum possible entanglement.
Most entanglement measures assign a value of 1 to a Bell state,
and 0 to all separable states.

For our purposes, it is convenient to use the tangle $\tau$
(squared concurrence)
as our entanglement measure when considering two or three qubits.
The tangle is an entanglement monotone from which the entanglement
of formation can be calculated \cite{wootters97a,coffman99a}.
For a pure state of two qubits, $\tau = 4\text{det}|\rho_A|$, where
$\rho_A$ is the reduced density matrix obtained when qubit $B$ has been
traced over (or vice versa permuting $A$ and $B$).
For a mixed state $\rho$ of two qubits, the concurrence $C=\tau^{1/2}$
is given \cite{wootters97a} by
\begin{equation}
C = \text{max}(\lambda_1 - \lambda_2 - \lambda_3 - \lambda_4, 0),
\label{eq:cur}
\end{equation}
where the $\lambda_i$ are the square roots of the eigenvalues of
$\rho \tilde{\rho} = \rho\;\sigma_{y}^{A}\!\otimes\!\sigma_{y}^{B}
\rho^{*} \sigma_y^A\!\otimes\!\sigma_{y}^{B}$,
and $\rho^{*}$ denotes the complex conjugation of $\rho$
in the computational basis $\{|00\rangle, |01\rangle,|10\rangle,|11\rangle\}$.
The entanglement of formation 
$E_f = h(\frac{1}{2} + \frac{1}{2}\sqrt{1-C^2})$,
where $h(x)$ is the binary entropy function,
$h(x) = -x\log_2(x) - (1-x) \log_2(1-x)$.

A general pure state of two qubits can be written \cite{nemoto00a}
\begin{eqnarray}
|\Psi_2\rangle
&=& e^{i\chi_0}\cos\theta_0 |00\rangle \nonumber \\
&+& e^{i\chi_1}\sin\theta_0 \cos\theta_1 |01\rangle \nonumber \\
&+& e^{i\chi_2}\sin\theta_0 \sin\theta_1 \cos\theta_2 |10\rangle \nonumber \\
&+& e^{i\chi_3}\sin\theta_0 \sin\theta_1 \sin\theta_2 |11\rangle,
\label{eq:su2}
\end{eqnarray}
where $\chi_j$, $\theta_j$ are chosen uniformly according to
\begin{equation}
(2\pi)^{-4}\mathrm{d}(\sin\theta_0)^6 \mathrm{d}(\sin\theta_1)^4
\mathrm{d}(\sin\theta_2)^2
\mathrm{d}\chi_0 \mathrm{d}\chi_1 \mathrm{d}\chi_2 \mathrm{d}\chi_3,
\label{eq:haar2}
\end{equation}
the Haar measure, with $0 \le\chi_i < 2\pi$ and $0 \le \theta_i < \pi/2$.
(We include an extra overall random phase, $e^{i\chi_0}$,
to maintain consistency with SU($n$).)

\begin{figure}
    \begin{minipage}{\columnwidth}
    \begin{center}
        \resizebox{0.8\columnwidth}{!}{\includegraphics{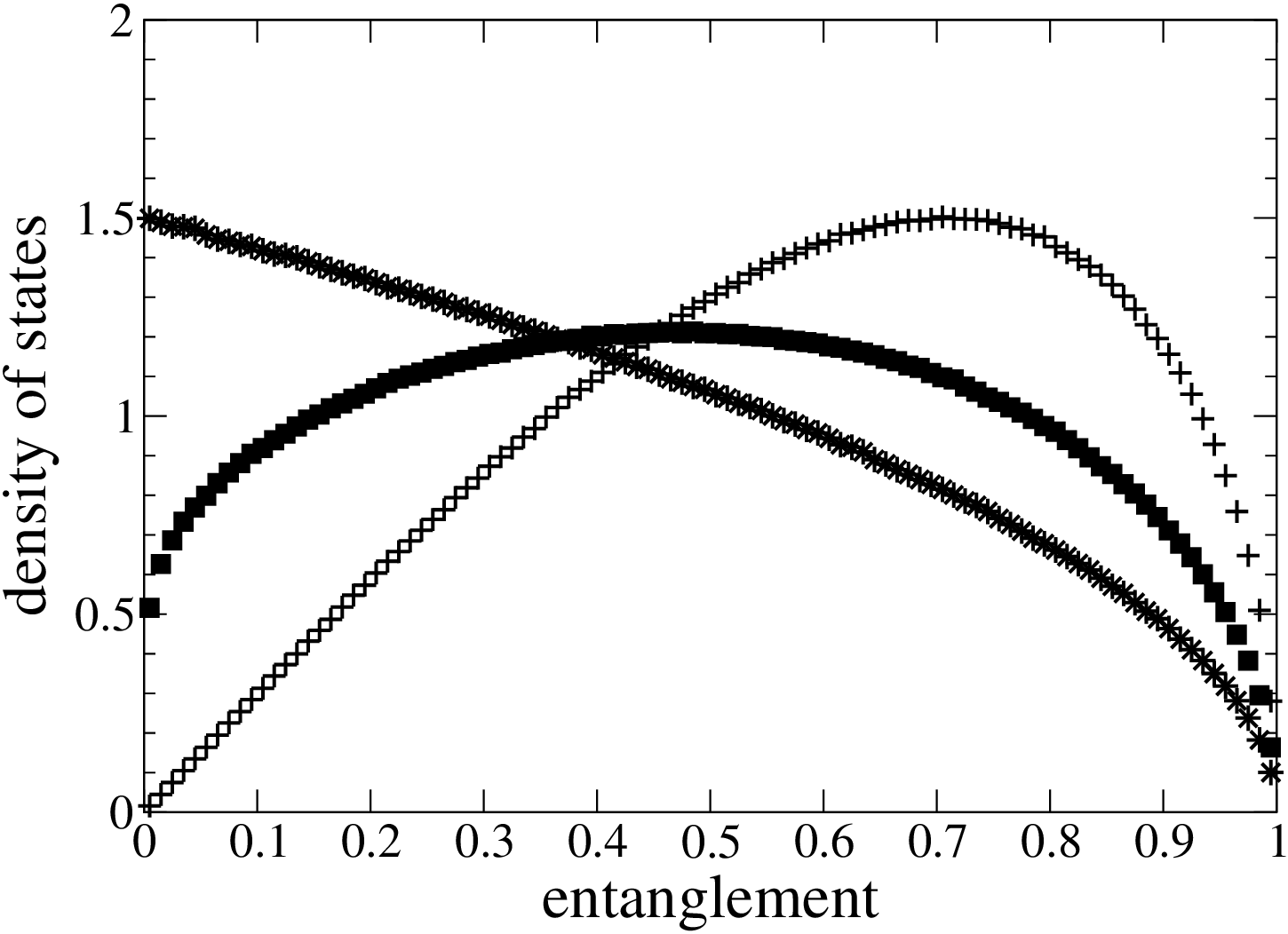}}
    \end{center}
    \end{minipage}
    \caption{Distribution of the tangle ($\ast$), concurrence ($+$)
    and entanglement of formation (\protect\rule[0ex]{1ex}{1ex})
    for two qubit pure states sampled uniformly over the Haar measure.}
    \label{fig:pdfs_2}
\end{figure}

Calculating $\tau = 4\text{det}|\rho_A|$ using Eq. (\ref{eq:su2})
and integrating over the Haar measure, Eq. (\ref{eq:haar2}),
gives $\langle\tau\rangle = 2/5$.
A randomly selected pure state of two qubits might thus be expected to have
0.4 tangle units of entanglement, and we have already noted that
states exist with the maximum ($=1$) and minimum ($=0$) amounts of
entanglement.  More informative is the distribution,
the density of states (over the Haar measure) with a given value of
the tangle.  Calculated numerically by sampling 30 million random pure
states, this distribution is shown in Fig. \ref{fig:pdfs_2}.
In tangle units, the distribution is broad, with many states
having little or no entanglement compared to only a few with high or
maximal entanglement.

However, it must be emphasized that the shape of the distribution is
heavily dependent on the choice of entanglement measure.  Compare in
Fig. \ref{fig:pdfs_2}
the distributions of the concurrence $C=\tau^{1/2}$ and the 
entanglement of formation (first shown in \cite{zyczkowski99a}).
The corresponding average values (calculated numerically and in
agreement with \cite{zyczkowski99a})
are $\langle C \rangle = 0.5890\pm0.0002$
and $\langle E_f \rangle = 0.4808\pm0.0002$.
Given the plethora of entanglement measures in current use,
it is important to only compare like with like.
Also, it is interesting to note that, though close,
none of these average entanglement
values are exactly equal to 1/2, which would be the na\"{\i}ve guess.

For three qubits, it is possible to define the 3-tangle $\tau_3$
for a pure state \cite{coffman99a}, giving a measure of the purely three-way
entanglement in the system.  A value for the tangle between each of
the three possible pairs of qubits can also be calculated,
$\tau_{AB}$, $\tau_{AC}$ and $\tau_{BC}$, using Eq. (\ref{eq:cur}).
These satisfy,
\begin{equation}
\tau_3 = \tau_A - \tau_{AB} - \tau_{AC}
\label{eq:tau3}
\end{equation}
where $\tau_A = 4\text{det}|\rho_A|$ as before, except now both
qubits B and C have been traced out to leave the partial density matrix
$\rho_A$.  Equation (\ref{eq:tau3}) holds for any permutation of
$A$, $B$ and $C$.
For the GHZ state $(|000\rangle + |111\rangle)/\sqrt{2}$,
which has the maximum possible 3-tangle,
$\tau_3 = 1$ and $\tau_{AB} = 0$, while for the W state
$(|001\rangle + |010\rangle + |100\rangle)/\sqrt{3}$,
$\tau_3 = 0$ and $\tau_{AB} = 4/9$ for each pair,
the maximum possible amount of pairwise tangle in a three qubit state
\cite{koashi00a}.

As for two qubits, the average values for $\tau_3$ and $\tau_{AB}$ can be
calculated.  After analytically evaluating somewhat lengthy integrals,
they are found to be $\langle\tau_3\rangle = 1/3$
and $\langle\tau_{AB}\rangle = 1/6$.
Since $\tau_3$ and $\tau_{AB}$ satisfy Eq. (\ref{eq:tau3}), one
can consider the quantity
$\langle \tau_3 + \tau_{AB} + \tau_{AC} + \tau_{BC} \rangle = 5/6$,
the average total entanglement
(in tangle units) in a random pure state
of three qubits.  This gives an average entanglement of
$5/18 \simeq 0.27$ per qubit for three qubits,
compared with 1/5 per qubit for two qubits.
A GHZ state has a tangle of 1/3 per qubit and a W state 4/9.
O'Connor and Wootters \cite{wootters00a,oconnor00a}
considered rings and chains of $N$ qubits in a translationally
invariant state and determined the maximum possible nearest neigbour
entanglement to be at least $C=0.434467$ or $\tau = 0.18876157$ per qubit.

\begin{figure}
    \begin{minipage}{\columnwidth}
    \begin{center}
        \resizebox{0.8\columnwidth}{!}{\includegraphics{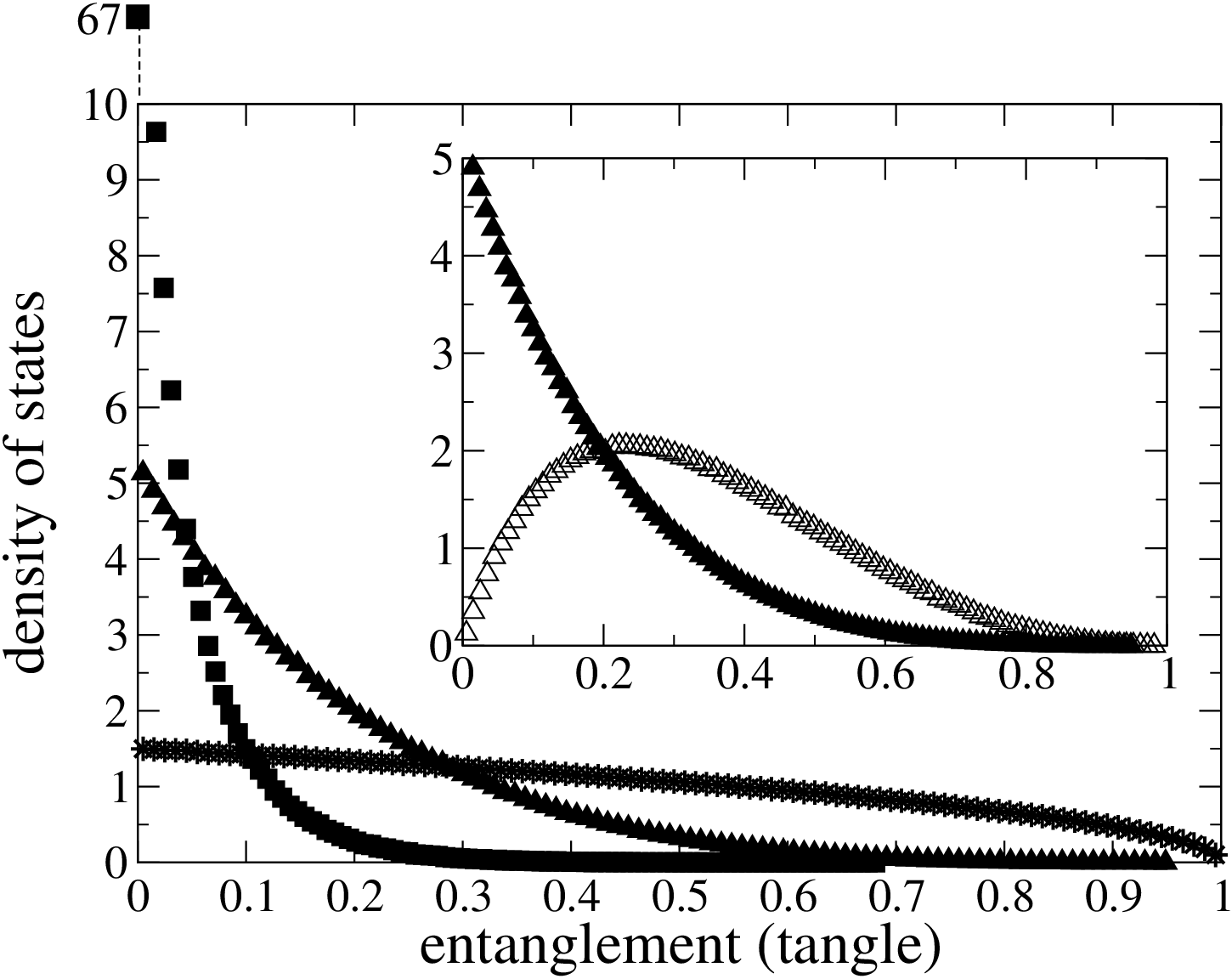}}
    \end{center}
    \end{minipage}
    \caption{ Distribution of the
    2-tangle for two ($\ast$), three (filled $\triangle$) and four
    (\protect\rule[0ex]{1ex}{1ex}) qubit pure states;
    inset: distribution of the 3-tangle ($\triangle$) and pairwise tangle
    (filled $\triangle$) for three qubit pure states,
    all sampled uniformly over the Haar measure.}
    \label{fig:tau_34}
\end{figure}

The distributions for $\tau_3$ and $\tau_{AB}$,
Fig. \ref{fig:tau_34} (inset), were calculated numerically by
samping a million pure three qubit states drawn randomly over the
Haar measure, the generalization of Eq. (\ref{eq:haar2}).
The 2-tangle is now much more concentrated near zero than for two qubits,
while the 3-tangle is broadly peaked around the average value.

These values for pairwise and three-way tangles would be more useful
if it was possible to continue in the same manner for four and more qubits.
Currently, there is no known analytical expression for the
3-tangle of a mixed state of three qubits, nor is it known whether 
expressions equivalent to Eq. (\ref{eq:tau3}) can be found beyond three qubits.
However, we can at least continue to evaluate pairwise entanglement for
larger pure states.
For four qubits the average pairwise entanglement evaluated numerically
from a million randomly sampled four qubit pure states 
is $0.03138\pm0.00006$ tangle units per pair.
Despite the low value for pairwise entanglement,
a typical four qubit pure state is still highly entangled overall
\cite{kempe01a}, as can be shown by considering the entropy of its subsystems.
For a pure state of N qubits, the entropy $S_N = 0$.  If the
system is split into two pieces, of $k$ qubits and $(N-k)$ qubits,
then each subsystem has the same entropy,
$S_k = S_{(N-k)} = \sum_i \lambda_i \log_2 \lambda_i$,
where the $\lambda_i$ are the eigenvalues of the reduced density matrix
$\rho_k$ obtained by tracing out the other $(N-k)$ qubits (or vice versa).
The entropy $S_k$ measures how entangled the two subsystems are
with each other.  For four qubit pure states, $\langle S_1 \rangle = 
\langle S_3 \rangle = 0.8661\dots$ and $\langle S_2 \rangle = 0.6653\dots$
normalised per qubit such that $0 \le S_k \le 1$.

The four qubit pairwise distribution, shown in Fig. \ref{fig:tau_34},
is now strongly peaked around zero entanglement.
A closer look at the numerical data
reveals a new feature not present in the distributions for two or three
qubit pure states: there is now a significant fraction of the pairs
(24\%) with zero entanglement, causing the outlying point
at 67 for the bin that includes zero.
This is not just zero to numerical accuracy.
The formula for the concurrence of a pair of qubits
in a mixed state $\rho$ is given by Eq. (\ref{eq:cur}).  Clearly, if
the values of the $\lambda_i$ are such that the maximum is zero by
a significant margin, then the result is effectively exact.
This trend continues for random pure states of five and six qubits, with
(evaluated numerically over samples of 100,000)
80\% and 99\% respectively of pairs having zero entanglement
in such states selected randomly over the Haar measure.
This is the probably $P_2$ of finding a chosen pair has zero entanglement.
The probability $P_s$ that the state as a whole has zero pairwise entanglement
in all possible pairs is smaller, but also fast
approaching $100\%$; by $N=7$ numerical sampling gives $P_s > 99.9\%$.
So we have a picture of typical pure states having less and less
pairwise entanglement as the number of qubits increases, with the
measure of states with some pairwise entanglement becoming essentially
zero beyond 7 qubits.  In other words, the fall off is not gradual,
but occurs sharply between $3 < N \lesssim 7$.

This immediately begs the question, what happens to the entanglement in
subsets of three or more qubits as N increases?
While it isn't known how to calculate
the 3-tangle for mixed states of three qubits, there is another
measure of entanglement that can be used to answer this
question.  The partial transpose of a density matrix expressed in the
standard basis is obtained by interchanging terms with a selected qubit
in the opposite state \cite{peres96a}.
Such a matrix, denoted $\rho^T$, can have
negative eigenvalues if the original density matrix ${\rho}$ is entangled.
If $\rho^T$ has only positive eigenvalues for any combination of transposed
qubits (up to half the total in $\rho$), then $\rho$ is said to be 
PPT (positive partial transpose), and in most cases, $\rho$ is separable.
The exceptions have bound entanglement \cite{horodecki97a,horodecky98a}, which
only occurs in systems with Hilbert space larger than $2\otimes2$.
Bound entangled states are known to be relatively
rare \cite{zyczkowski99a}, of finite measure but smaller than the measure
of separable states, which decreases exponentially in the number of qubits.
\begin{figure}
    \begin{minipage}{\columnwidth}
    \begin{center}
        \resizebox{0.8\columnwidth}{!}{\includegraphics{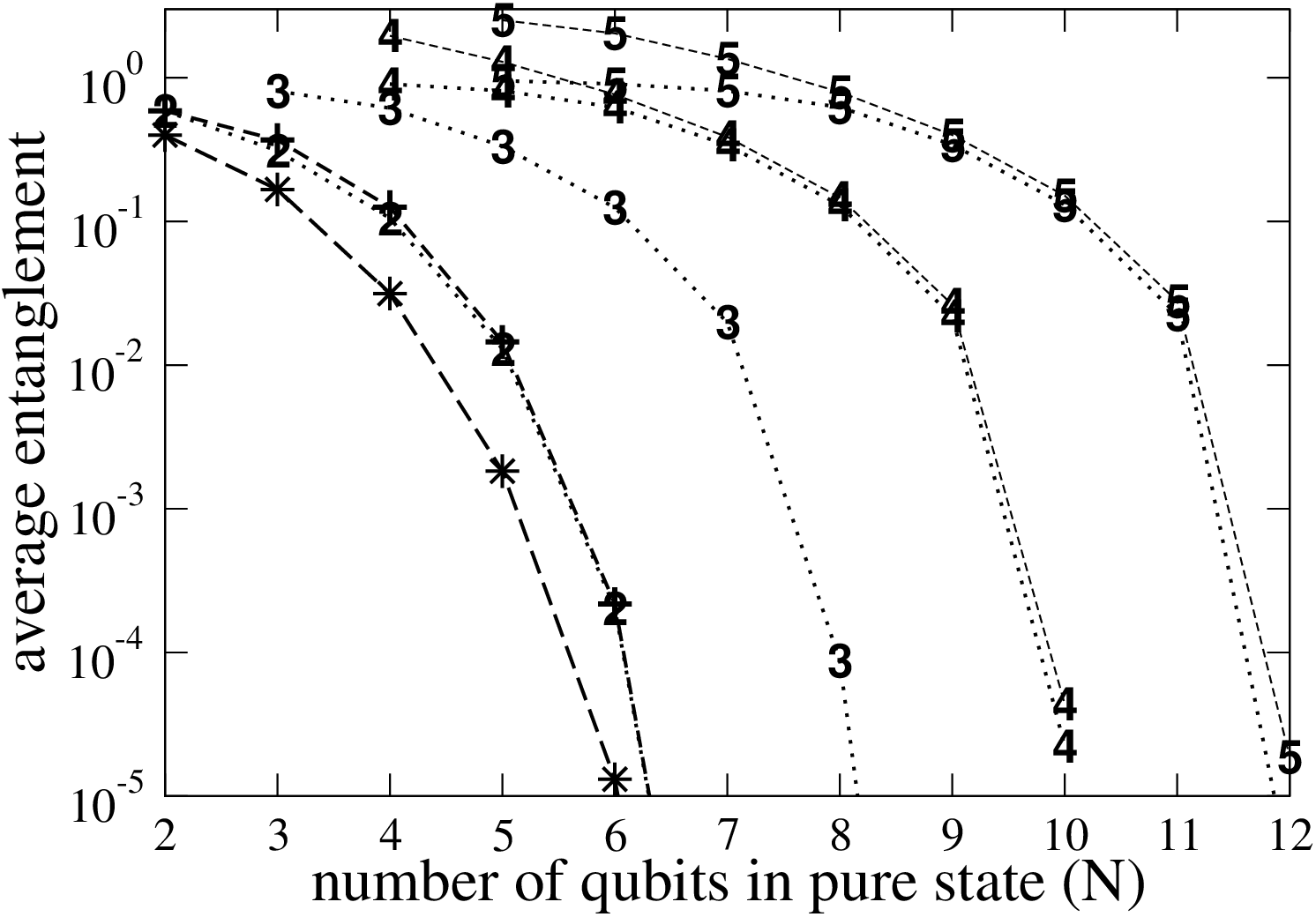}}
    \end{center}
    \end{minipage}
    \caption{Average negativity for subsets of $m$ qubits selected from random
    pure states sampled uniformly over the Haar measure.  The symbol gives $m$.
    For subsets of two qubits, the tangle ($\ast$) and concurrence($+$)
    are also shown.}
    \label{fig:neg_nq}
\end{figure}

A PPT test is sufficient to show that a three qubit subset has
no free entanglement, but there is also a quantitative measure
of entanglement that can be derived from $\rho^T$, the negativity.
Following \.{Z}yczkowski \cite{zyczkowski98a} we define the negativity
as twice the absolute value of the sum of the negative eigenvalues of $\rho^T$;
Vidal and Werner \cite{vidal01a} recently proved it is an entanglement monotone.
Figure \ref{fig:neg_nq} shows how the average values compare with the tangle
and concurrence,
two curves are shown for subsets of four and five qubits because there are different
values for transposing one qubit, or two qubits at once.
A log-linear plot has been used to show that all the entanglement measures
fall faster than exponentially as the number of qubits in the pure state
increases.

\begin{figure}
    \begin{minipage}{\columnwidth}
    \begin{center}
        \resizebox{0.8\columnwidth}{!}{\includegraphics{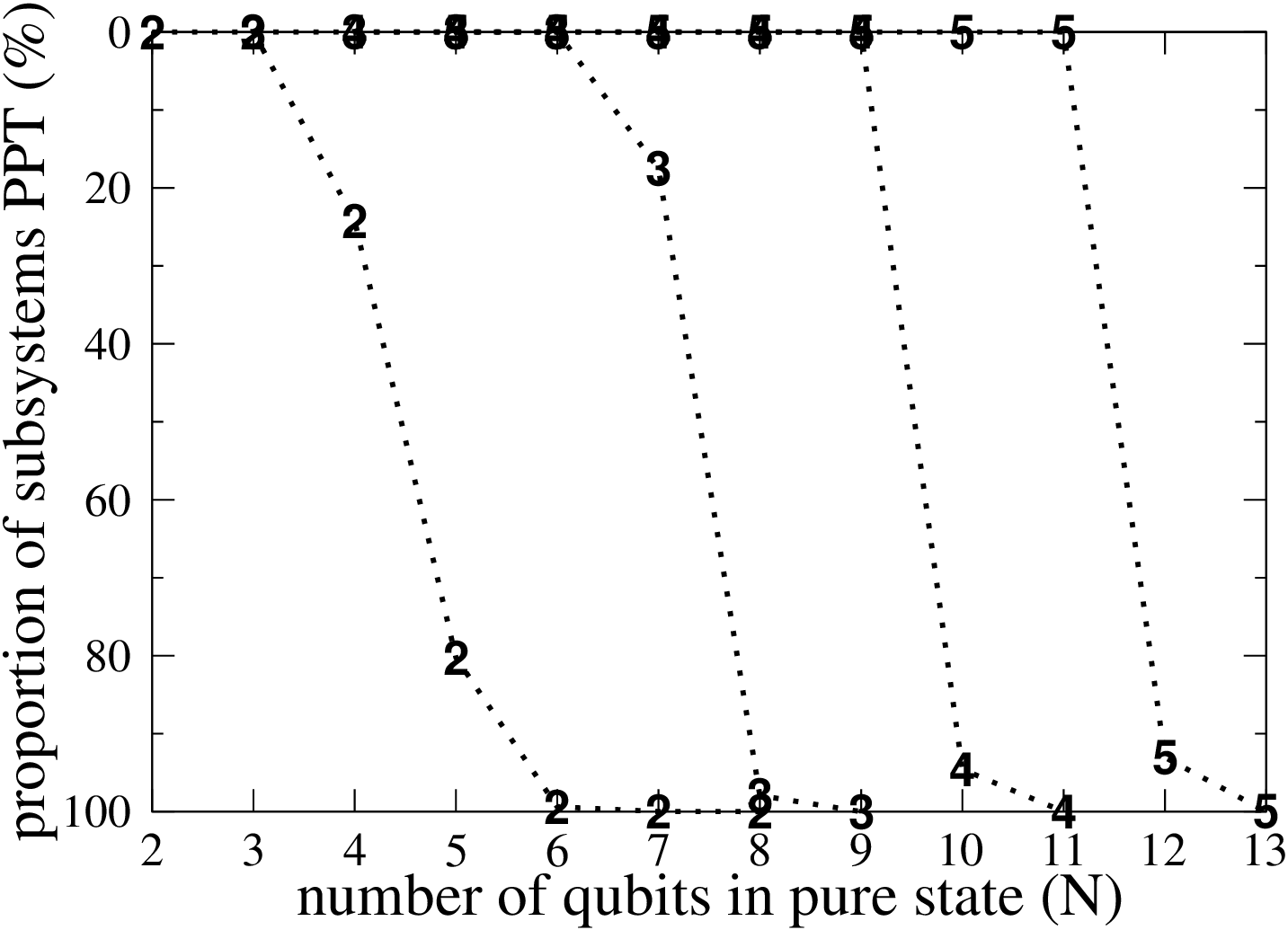}}
    \end{center}
    \end{minipage}
    \caption{Percentage of subsets of $m$ qubits with no (free) entanglement
	in random pure states sampled uniformly over the Haar measure,
	where $m$ is the symbol.}
    \label{fig:ppt_nq}
\end{figure}

Note, however, that the negativity does not satisfy any convenient
quantitative relation like Eq. (\ref{eq:tau3}) for the tangle.
It is thus not possible to identify finite amounts of three-way entanglement 
unless you also know that there is zero pairwise entanglement in the pure state.
Conversely, if the negativity of all subsets of three qubits
is zero, there is still the possibility of bound entanglement.
Nonetheless, since bound entanglement is a relatively rare phenomenon,
if we are willing to neglect it in order to obtain the shape of the
larger picture over all pure states, we can use the negativity to
identify states that have approximately zero entanglement in subsets of
three, four and five qubits, Fig. \ref{fig:ppt_nq}.
In each case, it drops sharply to zero between $N \simeq 2m$ and
$N=2m + 3$, based on numerical analysis of
random pure states sampled uniformly over the Haar measure for $N \le 13$ and
$m \le 6$.

\begin{figure}
    \begin{minipage}{\columnwidth}
    \begin{center}
        \resizebox{0.8\columnwidth}{!}{\includegraphics{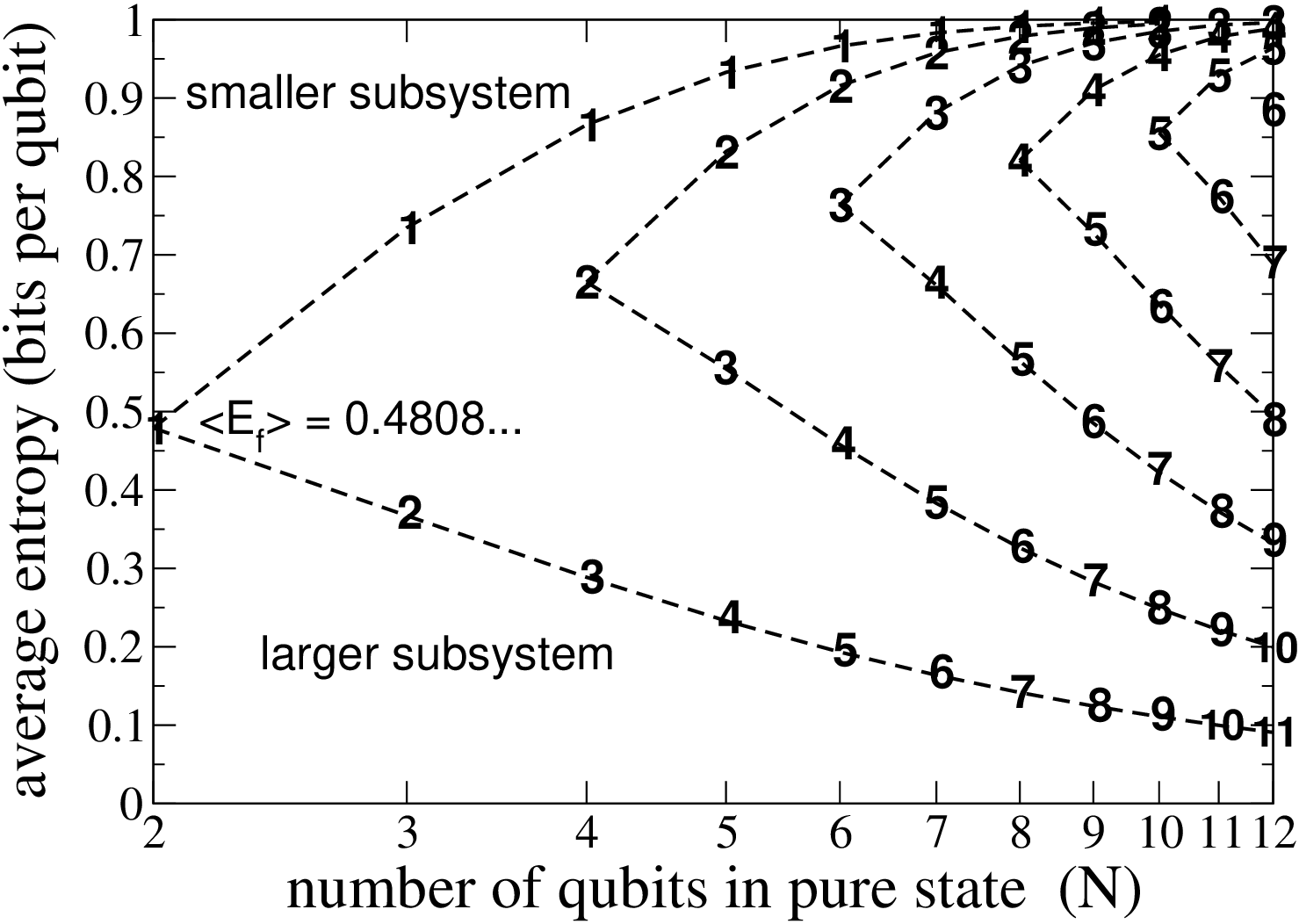}}
    \end{center}
    \end{minipage}
    \caption{Average entropy of subsystems plotted per qubit for random pure
    states sampled uniformly over the Haar measure.  Symbols give
    subsystem size, $k$ and $(N-k)$.
    For $N=2$, the average entropy corresponds to
    $\langle E_f\rangle$ for two qubits, as shown.}
    \label{fig:sent_nq}
\end{figure}

This result can be explained at least qualitatively as follows.
Page \cite{page93a} conjectured (later proved \cite{foong94a,sen96a})
that the average entropy of such a subsystem is (in the 
notation of this paper)
\begin{eqnarray}
\langle S_k\rangle\,\ln(2) &=& \sum_{j=2^{N-k}+1}^{2^N} \frac{1}{j} \;
                - \frac{2^k-1}{2^{N-k+1}} \nonumber \\
           &\simeq& k\ln(2) - \frac{1}{2^{N-2k+1}}
\label{eq:sent}
\end{eqnarray}
where the approximate version is valid for $1 \ll k \le N/2$.
These average entropies can also be calculated numerically in
the same manner as we calculated the tangle and concurrence,
by sampling random pure states.  Comparison of these values with the
theoretical formula provides a useful test of the accuracy of the
numerical method, and agreement was generally better than 3 significant
digits, even for relatively small sample sizes.
A useful way of presenting $\langle S_k \rangle$ is shown in
Fig. \ref{fig:sent_nq}, where it is plotted
per qubit for both the larger and smaller of the two subsystems.
Equation (\ref{eq:sent}) implies that the entropy of the smaller subsystem
is nearly maximal, which in turn implies that the subsystem is nearly
maximally mixed, i.e. not entangled.  In Fig. \ref{fig:sent_nq} this
is shown by the lines for the smaller subsystem asymptoting to 1 as
the difference between $k$ and $N$ becomes larger.
There is a set of unentangled states of finite measure \cite{zyczkowski99a}
surrounding the maximally mixed state ($\rho = \bm{1}$).
As $N$ increases and the smaller subsystem becomes more mixed,
it enters this unentangled region for finite $N$, thus explaining the
rapid transition to zero entanglement shown in Fig. \ref{fig:ppt_nq}.


In conclusion, we have shown that in randomly chosen pure states of $N$ qubits,
there is a cut-off, $m$, below which subsets of qubits taken from the pure
state by partial tracing over the remainder can be expected to be
PPT (i.e. no free entanglement).
For $N \le 13$, the limited range accessible to our numerical studies,
$m \gtrsim (N-3)/2$.
There are already experimental situations (such as ion traps) 
with more than six distinguishable quantum particles in which such
entangled states might be generated.  Only looking for pairwise entanglement
in such systems may give misleading results.

\begin{acknowledgments}
This work was funded in part by
the European project EQUIP(IST-1999-11053) and QUIPROCONE (IST-1999-29064).
VK is funded by the UK Engineering and Physical Sciences
Research Council grant number GR/N2507701.
KN is funded in part under project QUICOV as part of the
IST-FET-QJPC programme.
We would like to thank Seth Lloyd for drawing our attention to 
Refs. \cite{page93a,foong94a,sen96a}, and Martin Plenio, Karol \.Zyczkowski
and Jens Eisert for useful discussions.
\end{acknowledgments}


\bibliography{ent}

\begin{thebibliography}{10}
\expandafter\ifx\csname bibnamefont\endcsname\relax
  \def\bibnamefont#1{#1}\fi
\expandafter\ifx\csname bibfnamefont\endcsname\relax
  \def\bibfnamefont#1{#1}\fi
\expandafter\ifx\csname url\endcsname\relax
  \def\url#1{\texttt{#1}}\fi
\expandafter\ifx\csname urlprefix\endcsname\relax\def\urlprefix{URL }\fi
\providecommand{\bibinfo}[2]{#2}
\providecommand{\eprint}[2][]{\url{#2}}

\bibitem{divincenzo95a}
\bibinfo{author}{\bibfnamefont{D.~P.} \bibnamefont{DiVincenzo}},
  \bibinfo{journal}{Science} \textbf{\bibinfo{volume}{270}},
  \bibinfo{pages}{255} (\bibinfo{year}{1995}).

\bibitem{vedral98b}
\bibinfo{author}{\bibfnamefont{V.}~\bibnamefont{Vedral}} \bibnamefont{and}
  \bibinfo{author}{\bibfnamefont{M.~B.} \bibnamefont{Plenio}},
  \bibinfo{journal}{Prog. Quant. Electron.} \textbf{\bibinfo{volume}{22}},
  \bibinfo{pages}{1} (\bibinfo{year}{1998}).

\bibitem{schumacher96a}
\bibinfo{author}{\bibfnamefont{B.}~\bibnamefont{Schumacher}},
  \bibinfo{journal}{Phys. Rev. A} \textbf{\bibinfo{volume}{54}},
  \bibinfo{pages}{2614} (\bibinfo{year}{1996}).

\bibitem{tittel00a}
\bibinfo{author}{\bibfnamefont{W.}~\bibnamefont{Tittel}},
  \bibinfo{author}{\bibfnamefont{J.}~\bibnamefont{Brendel}},
  \bibinfo{author}{\bibfnamefont{H.}~\bibnamefont{Zbinden}}, \bibnamefont{and}
  \bibinfo{author}{\bibfnamefont{N.}~\bibnamefont{Gisin}},
  \bibinfo{journal}{Phys. Rev. Lett.} \textbf{\bibinfo{volume}{84}},
  \bibinfo{pages}{4737} (\bibinfo{year}{2000}).

\bibitem{jennewein00a}
\bibinfo{author}{\bibfnamefont{T.}~\bibnamefont{Jennewein}},
  \bibinfo{author}{\bibfnamefont{C.}~\bibnamefont{Simon}},
  \bibinfo{author}{\bibfnamefont{G.}~\bibnamefont{Weihs}},
  \bibinfo{author}{\bibfnamefont{H.}~\bibnamefont{Weinfurter}},
  \bibnamefont{and}
  \bibinfo{author}{\bibfnamefont{A.}~\bibnamefont{Zeilinger}},
  \bibinfo{journal}{Phys. Rev. Lett.} \textbf{\bibinfo{volume}{84}},
  \bibinfo{pages}{4729} (\bibinfo{year}{2000}).

\bibitem{naik00a}
\bibinfo{author}{\bibfnamefont{D.~S.} \bibnamefont{Naik}},
  \bibinfo{author}{\bibfnamefont{C.~G.} \bibnamefont{Peterson}},
  \bibinfo{author}{\bibfnamefont{A.~G.} \bibnamefont{White}},
  \bibinfo{author}{\bibfnamefont{A.~J.} \bibnamefont{Berglund}},
  \bibnamefont{and} \bibinfo{author}{\bibfnamefont{P.~G.} \bibnamefont{Kwiat}},
  \bibinfo{journal}{Phys. Rev. Lett.} \textbf{\bibinfo{volume}{84}},
  \bibinfo{pages}{4733} (\bibinfo{year}{2000}).

\bibitem{ekert91a}
\bibinfo{author}{\bibfnamefont{A.~K.} \bibnamefont{Ekert}},
  \bibinfo{journal}{Phys. Rev. Lett.} \textbf{\bibinfo{volume}{67}},
  \bibinfo{pages}{661} (\bibinfo{year}{1991}).

\bibitem{boschi98a}
\bibinfo{author}{\bibfnamefont{D.}~\bibnamefont{Boschi}},
  \bibinfo{author}{\bibfnamefont{S.}~\bibnamefont{Branca}},
  \bibinfo{author}{\bibfnamefont{F.}~\bibnamefont{{De Martini}}},
  \bibinfo{author}{\bibfnamefont{L.}~\bibnamefont{Hardy}}, \bibnamefont{and}
  \bibinfo{author}{\bibfnamefont{S.}~\bibnamefont{Popescu}},
  \bibinfo{journal}{Phys. Rev. Lett.} \textbf{\bibinfo{volume}{80}},
  \bibinfo{pages}{1121} (\bibinfo{year}{1998}).

\bibitem{bennett93a}
\bibinfo{author}{\bibfnamefont{C.~H.} \bibnamefont{Bennett}},
  \bibinfo{author}{\bibfnamefont{G.}~\bibnamefont{Brassard}},
  \bibinfo{author}{\bibfnamefont{C.}~\bibnamefont{Cr\'{e}peau}},
  \bibinfo{author}{\bibfnamefont{R.}~\bibnamefont{Jozsa}},
  \bibinfo{author}{\bibfnamefont{A.}~\bibnamefont{Peres}}, \bibnamefont{and}
  \bibinfo{author}{\bibfnamefont{W.~K.} \bibnamefont{Wootters}},
  \bibinfo{journal}{Phys. Rev. Lett.} \textbf{\bibinfo{volume}{70}},
  \bibinfo{pages}{1895} (\bibinfo{year}{1993}).

\bibitem{bouwmeester97a}
\bibinfo{author}{\bibfnamefont{D.}~\bibnamefont{Bouwmeester}},
  \bibinfo{author}{\bibfnamefont{J.~W.} \bibnamefont{Pan}},
  \bibinfo{author}{\bibfnamefont{K.}~\bibnamefont{Mattle}},
  \bibinfo{author}{\bibfnamefont{M.}~\bibnamefont{Eibl}},
  \bibinfo{author}{\bibfnamefont{H.}~\bibnamefont{Weinfurter}},
  \bibnamefont{and}
  \bibinfo{author}{\bibfnamefont{A.}~\bibnamefont{Zeilinger}},
  \bibinfo{journal}{Nature} \textbf{\bibinfo{volume}{390}},
  \bibinfo{pages}{575} (\bibinfo{year}{1997}).

\bibitem{wootters97a}
\bibinfo{author}{\bibfnamefont{W.~K.} \bibnamefont{Wootters}},
  \bibinfo{journal}{Phys. Rev. Lett.} \textbf{\bibinfo{volume}{80}},
  \bibinfo{pages}{2245} (\bibinfo{year}{1998}), \eprint{quant-ph/9709029}.

\bibitem{zyczkowski98a}
\bibinfo{author}{\bibfnamefont{K.}~\bibnamefont{{\.{Z}}yczkowski}},
  \bibinfo{author}{\bibfnamefont{P.}~\bibnamefont{Horodecki}},
  \bibinfo{author}{\bibfnamefont{A.}~\bibnamefont{Sanpera}}, \bibnamefont{and}
  \bibinfo{author}{\bibfnamefont{M.}~\bibnamefont{Lewenstein}},
  \bibinfo{journal}{Phys. Rev. A} \textbf{\bibinfo{volume}{58}},
  \bibinfo{pages}{883} (\bibinfo{year}{1998}), \eprint{quant-ph/9804024}.

\bibitem{wootters00a}
\bibinfo{author}{\bibfnamefont{W.~K.} \bibnamefont{Wootters}},
  \emph{\bibinfo{title}{Entangled chains}} (\bibinfo{year}{2000}),
  \eprint{quant-ph/0001114}.

\bibitem{oconnor00a}
\bibinfo{author}{\bibfnamefont{K.~M.} \bibnamefont{O'Connor}} \bibnamefont{and}
  \bibinfo{author}{\bibfnamefont{W.~K.} \bibnamefont{Wootters}},
  \bibinfo{journal}{Phys. Rev. A} \textbf{\bibinfo{volume}{63}},
  \bibinfo{pages}{052302} (\bibinfo{year}{2001}), \eprint{quant-ph/0009041}.

\bibitem{koashi00a}
\bibinfo{author}{\bibfnamefont{M.}~\bibnamefont{Koashi}},
  \bibinfo{author}{\bibfnamefont{V.}~\bibnamefont{{Bu\~{z}ek}}},
  \bibnamefont{and} \bibinfo{author}{\bibfnamefont{N.}~\bibnamefont{Imoto}},
  \bibinfo{journal}{Phys. Rev. A} \textbf{\bibinfo{volume}{62}},
  \bibinfo{pages}{050302(R)} (\bibinfo{year}{2000}), \eprint{quant-ph/0007086}.

\bibitem{gunlycke01a}
\bibinfo{author}{\bibfnamefont{D.}~\bibnamefont{Gunlycke}},
  \bibinfo{author}{\bibfnamefont{V.~M.} \bibnamefont{Kendon}},
  \bibinfo{author}{\bibfnamefont{V.}~\bibnamefont{Vedral}}, \bibnamefont{and}
  \bibinfo{author}{\bibfnamefont{S.}~\bibnamefont{Bose}},
  \bibinfo{journal}{Phys. Rev. A} \textbf{\bibinfo{volume}{64}},
  \bibinfo{pages}{042302} (\bibinfo{year}{2001}), \eprint{quant-ph/0102137}.

\bibitem{raussendorf01a}
\bibinfo{author}{\bibfnamefont{R.}~\bibnamefont{Raussendorf}} \bibnamefont{and}
  \bibinfo{author}{\bibfnamefont{H.~J.} \bibnamefont{Briegel}},
  \bibinfo{journal}{Phys. Rev. Lett.} \textbf{\bibinfo{volume}{86}},
  \bibinfo{pages}{5188} (\bibinfo{year}{2001}).

\bibitem{bennett96b}
\bibinfo{author}{\bibfnamefont{C.~H.} \bibnamefont{Bennett}},
  \bibinfo{author}{\bibfnamefont{D.~P.} \bibnamefont{DiVincenzo}},
  \bibinfo{author}{\bibfnamefont{J.~A.} \bibnamefont{Smolin}},
  \bibnamefont{and} \bibinfo{author}{\bibfnamefont{W.~K.}
  \bibnamefont{Wootters}}, \bibinfo{journal}{Phys. Rev. A}
  \textbf{\bibinfo{volume}{54}}, \bibinfo{pages}{3824} (\bibinfo{year}{1996}).

\bibitem{vedral97b}
\bibinfo{author}{\bibfnamefont{V.}~\bibnamefont{Vedral}},
  \bibinfo{author}{\bibfnamefont{M.~B.} \bibnamefont{Plenio}},
  \bibinfo{author}{\bibfnamefont{K.}~\bibnamefont{Jacobs}}, \bibnamefont{and}
  \bibinfo{author}{\bibfnamefont{P.~L.} \bibnamefont{Knight}},
  \bibinfo{journal}{Phys.Rev. A} \textbf{\bibinfo{volume}{56}},
  \bibinfo{pages}{4452} (\bibinfo{year}{1997}), \eprint{quant-ph/9703025}.

\bibitem{coffman99a}
\bibinfo{author}{\bibfnamefont{V.}~\bibnamefont{Coffman}},
  \bibinfo{author}{\bibfnamefont{J.}~\bibnamefont{Kundu}}, \bibnamefont{and}
  \bibinfo{author}{\bibfnamefont{W.~K.} \bibnamefont{Wootters}},
  \bibinfo{journal}{Phys. Rev. A} \textbf{\bibinfo{volume}{61}},
  \bibinfo{pages}{052306} (\bibinfo{year}{2000}), \eprint{quant-ph/9907047}.

\bibitem{nemoto00a}
\bibinfo{author}{\bibfnamefont{K.}~\bibnamefont{Nemoto}}, \bibinfo{journal}{J.
  Phys. A: Math. Gen.} \textbf{\bibinfo{volume}{33}}, \bibinfo{pages}{3493}
  (\bibinfo{year}{2000}), \eprint{quant-ph/0004087}.

\bibitem{zyczkowski99a}
\bibinfo{author}{\bibfnamefont{K.}~\bibnamefont{{\.{Z}}yczkowski}},
  \bibinfo{journal}{Phys. Rev. A} \textbf{\bibinfo{volume}{60}},
  \bibinfo{pages}{3496} (\bibinfo{year}{1999}), \eprint{quant-ph/9902050}.

\bibitem{kempe01a}
\bibinfo{author}{\bibfnamefont{M.~A.} \bibnamefont{Nielsen}} \bibnamefont{and}
  \bibinfo{author}{\bibfnamefont{J.}~\bibnamefont{Kempe}},
  \bibinfo{journal}{Phys. Rev. Lett.}
  \textbf{\bibinfo{volume}{86}}(\bibinfo{number}{22}), \bibinfo{pages}{5184}
  (\bibinfo{year}{2001}), \eprint{quant-ph/0011117}.

\bibitem{peres96a}
\bibinfo{author}{\bibfnamefont{A.}~\bibnamefont{Peres}},
  \bibinfo{journal}{Phys. Rev. Lett.} \textbf{\bibinfo{volume}{77}},
  \bibinfo{pages}{1413} (\bibinfo{year}{1996}), \eprint{quant-ph/9604005}.

\bibitem{horodecki97a}
\bibinfo{author}{\bibfnamefont{P.}~\bibnamefont{Horodecki}},
  \bibinfo{journal}{Phys. Lett. A} \textbf{\bibinfo{volume}{232}},
  \bibinfo{pages}{333} (\bibinfo{year}{1997}), \eprint{quant-ph/9703004}.

\bibitem{horodecky98a}
\bibinfo{author}{\bibfnamefont{M.}~\bibnamefont{Horodecki}},
  \bibinfo{author}{\bibfnamefont{P.}~\bibnamefont{Horodecki}},
  \bibnamefont{and}
  \bibinfo{author}{\bibfnamefont{R.}~\bibnamefont{Horodecki}},
  \bibinfo{journal}{Phys. Rev. Lett.} \textbf{\bibinfo{volume}{80}},
  \bibinfo{pages}{5239} (\bibinfo{year}{1998}), \eprint{quant-ph/9801069}.

\bibitem{vidal01a}
\bibinfo{author}{\bibfnamefont{G.}~\bibnamefont{Vidal}} \bibnamefont{and}
  \bibinfo{author}{\bibfnamefont{R.~F.} \bibnamefont{Werner}},
  \emph{\bibinfo{title}{A computable measure of entanglement}}
  (\bibinfo{year}{2001}), \eprint{quant-ph/0102117}.

\bibitem{page93a}
\bibinfo{author}{\bibfnamefont{D.~N.} \bibnamefont{Page}},
  \bibinfo{journal}{Phys. Rev. Lett} \textbf{\bibinfo{volume}{71}},
  \bibinfo{pages}{1291} (\bibinfo{year}{1993}), \eprint{gr-qc/9305007}.

\bibitem{foong94a}
\bibinfo{author}{\bibfnamefont{S.~K.} \bibnamefont{Foong}} \bibnamefont{and}
  \bibinfo{author}{\bibfnamefont{S.}~\bibnamefont{Kanno}},
  \bibinfo{journal}{Phys. Rev. Lett.} \textbf{\bibinfo{volume}{72}},
  \bibinfo{pages}{1148} (\bibinfo{year}{1994}).

\bibitem{sen96a}
\bibinfo{author}{\bibfnamefont{S.}~\bibnamefont{Sen}}, \bibinfo{journal}{Phys.
  Rev. Lett.} \textbf{\bibinfo{volume}{77}}, \bibinfo{pages}{1}
  (\bibinfo{year}{1996}), \eprint{hep-th/9601132}.

\end{thebibliography}


\end{document}